\documentclass[aps,prd,a4paper,nofootinbib,
preprintnumbers,superscriptaddress,showpacs,showkeys,twocolumn]{revtex4-1}

\usepackage{amsmath}
\usepackage{amssymb}
\usepackage{bm}
\usepackage{graphicx}
\usepackage{color}
\usepackage[normalem]{ulem}  

\newcommand{\be}{\begin{equation}}
\newcommand{\ee}{\end{equation}}
\newcommand{\ba}{\begin{eqnarray}}
\newcommand{\ea}{\end{eqnarray}}

\renewcommand\sout{\bgroup \color{blue} \ULdepth=-.5ex \ULset} 
 
\begin{document}

\title{The influence of electromagnetic fields on the generation of the directed and elliptic flows of heavy quark in relativistic heavy-ion collisions}

\author{Santosh K. Das}
\affiliation{School of Physical Sciences, Indian Institute of Technology Goa, Ponda-403401, Goa, India}
\affiliation{Helmholtz Research Academy Hesse for FAIR (HFHF), GSI Helmholtz Center for Heavy Ion Research, Campus Frankfurt, 60438 Frankfurt, Germany}

\author{Olga Soloveva}
\affiliation{GSI Helmholtzzentrum f\"ur Schwerionenforschung GmbH,Planckstrasse 1, D-64291 Darmstadt, Germany}

\author{Taesoo Song}
\affiliation{GSI Helmholtzzentrum f\"ur Schwerionenforschung GmbH,Planckstrasse 1, D-64291 Darmstadt, Germany}

\author{Elena Bratkovskaya}
\affiliation{GSI Helmholtzzentrum f\"ur Schwerionenforschung GmbH,Planckstrasse 1, D-64291 Darmstadt, Germany}
\affiliation{Institut f\"ur Theoretische Physik, Johann Wolfgang Goethe-Universit\"at,Max-von-Laue-Str.\ 1, D-60438 Frankfurt am Main, Germany}
\affiliation{Helmholtz Research Academy Hesse for FAIR (HFHF), GSI Helmholtz Center for Heavy Ion Research, Campus Frankfurt, 60438 Frankfurt, Germany}


\begin{abstract}
We study the impact of self-generated electromagnetic fields (EMF) on the charm quarks momentum evolution  in the partonic and hadronic medium created in heavy-ion collisions at RHIC energy within the Parton-Hadron-String Dynamics (PHSD) off-shell transport approach. In the quark-gluon plasma (QGP) phase, the charm quark interacts with the off-shell partons whose mass and widths are given by the Dynamical Quasi-Particle Model (DQPM), which can reproduce the lattice QCD thermodynamics. The background electromagnetic fields are computed dynamically within the PHSD considering  both the spectators and participants protons as well as newly produced charged hadrons, quarks, and antiquarks, which reflects naturally the electric conductivity $\sigma_{el}$ of the medium.  We study the directed and elliptic flow of the $D$ mesons in the presence of the electromagnetic fields. 
We find that electromagnetically induced splitting in the $D$ meson $v_1$  through $D^0$ and $\overline{D}^0$  mesons  is consistent with the experimental data.  Furthermore, we notice that the splitting in the heavy quark $v_1$ as a function of $p_T$ is more prominent as a probe of the produced electromagnetic fields. 
However, we find only a small impact of electromagnetic fields on the heavy quark elliptic flow $v_2$.
\end{abstract}



\maketitle

\section{Introduction}
According to the fundamental theory of strong interactions, the quantum chromodynamics (QCD), at high temperature and density,  nuclear matter changes its phase;  hadrons dissolve to interacting quarks and gluons - the quark-gluon plasma (QGP)~\cite{Shuryak:2004cy, Jacak:2012dx}.  Relativistic heavy-ion collisions at the Relativistic Heavy-Ion Collider (RHIC) and the Large Hadron Collider (LHC)  are the experiments to realize such extreme conditions of temperature and density. 

To characterize the QGP, penetrating and well-calibrated probes are extremely important  and in this context, heavy quarks (HQs), mainly charm (c) and bottom (b) quarks, play a crucial role~\cite{Prino:2016cni,Andronic:2015wma,Rapp:2018qla,Aarts:2016hap,Cao:2018ews,Dong:2019unq,Xu:2018gux,Das:2024vac}.
Due to their larger mass $(M)$, compared to the temperature $(T)$, they are expected to be created early at RHIC and LHC energies, HQ thermal production in the QGP can be neglected because of Boltzmann suppression ( $\sim e^{-M/T}$); this ensures  nearly exact flavor conservation during the evolution of the QGP. Since the thermalization time of heavy quarks is delayed relative to the light partons of the bulk medium by a factor of order $\sim M/T$;  heavy flavour particles are not expected to be fully thermalized and therefore preserve a memory of their interaction history of both the initial stage and the subsequent evolution into the QGP phase.
One of the prime goals of hard probe research is to quantify the spatial diffusion constant, $D_s$, of the heavy quark,  a measure of the interaction strength of heavy quarks with the bulk medium. The collective properties of 
the $D$ meson have been measured both at RHIC and LHC energies through the nuclear suppression factor, $R_{AA}$ and elliptic flow, $v_2$. Several attempts ~\cite{vanHees:2005wb,vanHees:2007me,Gossiaux:2008jv, Gossiaux:2009mk,Das:2010tj,Alberico:2011zy,Lang:2012cx,Uphoff:2011ad,He:2012df,Uphoff:2012gb,Das:2013kea,Das:2015ana, Song:2015sfa,Nahrgang:2014vza,Song:2015ykw, Cao:2016gvr, Scardina:2017ipo,Xu:2017obm,Plumari:2017ntm,Katz:2019fkc}    have been made  to study both observables simultaneously which can constrain  the heavy quark diffusion coefficient.

In the recent past, it has been recognized that extremely intense electromagnetic fields~\cite{Skokov:2009qp, Voronyuk:2011jd} are produced in
non-central heavy-ion collisions mainly due to the motion of spectator charges. The strength of the 
initial magnetic field produced at the very early stage of the collisions at RHIC and LHC energies can be up to $eB\sim 5-50 \ m_{\pi}^2$, which is about a few orders of magnitude larger than that
expected to be produced at the surface of magnetars. Once the two spectator charges recede from each
other, the magnetic fields  decay, which in return generates an electric field. 
Heavy-ion collisions at RHIC and LHC energies provide a unique opportunity to study physics under extremely high electromagnetic fields. Heavy quarks are considered as a novel probe of this electromagnetic field considering the fact that they are produced early and being a non-equilibrium probe,  they will be able to retain the interaction history till its detection as heavy mesons in experiments. 
The directed flow $v_1$ of  the heavy mesons ~\cite{Das:2016cwd,Chatterjee:2018lsx,Oliva:2020doe,Oliva:2020mfr, Dubla:2020bdz, Sun:2020wkg, Beraudo:2021ont,Jiang:2022uoe,Sun:2023adv} is considered as a promising observable to characterize the generated electromagnetic fields. Heavy quark directed flow is predicted~\cite{Das:2016cwd} to be an order of magnitude larger than that of the light hadron directed flow. However, to disentangle other sources to the directed flow,  electromagnetically-induced splitting in the directed flow, between positively and negatively charged quarks, of charm and anti-charm quarks through $D^0$ and $\overline{D}^0$  mesons (also $D^-$ and $D^+$) is considered as a novel observable to scrutinize and quantify the initial electromagnetic field.

Recently both STAR~\cite{STAR:2019clv} and ALICE~\cite{ALICE:2019sgg} collaboration have measured the directed flow of  $D$ mesons at RHIC and LHC energies. Both collaborations obtained a finite directed flow which is an order of magnitude larger than that of the light hadrons. The measured splitting in the directed flow through $D^0$ and $\overline{D}^0$  mesons ($\Delta v_1=v_1(D^0)-v_1(\overline{D^0})$) at the highest RHIC center of mass energy of $\sqrt{s_{NN}}=200$ GeV fluctuates around zero and it is smaller than the current precision of the experimental measurements. On the other hand, the slope of the directed flow splitting measured at LHC energy is positive, and its magnitude is about three orders of magnitude larger than that of the light charged hadrons. 

So far, from the theoretical side, only very few attempts have been made to study the $D$ meson directed flow.  Many models obtained a finite but negative slope for the $D$ meson directed flow splitting both at RHIC and LHC energies \cite{Das:2016cwd,Oliva:2020doe,Chatterjee:2018lsx}. 
Recently, it has been shown that a positive slope of $D$ meson directed flow splitting at LHC energy can be obtained only if the magnetic field dominates over the electric field~\cite{Sun:2020wkg,Jiang:2022uoe}. 
However, it has been done in an ad-hoc way to  investigate a potential scenario yielding a positive slope.
The time evolution of the electromagnetic field plays an important role in setting the sign of the slope of a directed flow splitting. The models - currently employed to evaluate the electromagnetic field to compute heavy quark directed flow - involve several approximations.  Moreover, the electrical conductivity — an essential parameter for modeling the time evolution of the electromagnetic field in QCD matter — remains subject to significant uncertainties.

The goal of this study is to investigate the influence of electromagnetic fields on charm directed and elliptic flow  based on a consistent dynamical description of charm degrees-of-freedom (on quark and hadron levels) and their interactions in the partonic and hadronic medium, where electromagnetic fields are self-generated in dynamical way during the time evolution of heavy-ion collisions. Our study is based on  the Parton-Hadron-String Dynamics (PHSD)  \cite{Cassing:2008sv,Cassing:2008nn,Cassing:2009vt,Bratkovskaya:2011wp,Linnyk:2015rco,Moreau:2019vhw}, which is a microscopic covariant dynamical approach for strongly interacting systems formulated on the basis of off-shell Kadanoff-Baym equations, which describes the space-time evolution of the matter, starting from the initial hard collisions until kinetic freeze-out, produced in high energy heavy-ion collisions. 
The description of the QGP is done within the Dynamical QuasiParticle Model (DQPM) \cite{Cassing:2007nb,Cassing:2007yg,Cassing:2008nn,Linnyk:2015rco,Moreau:2019vhw,Soloveva:2020hpr} 
which is an effective field-theoretical model for the description of nonperturbative QCD phenomena and reproduces the lattice QCD thermodynamical results based on covariant propagators for  quarks/antiquarks and gluons that have a finite width in their spectral functions (imaginary parts of the propagators).  
The background electromagnetic fields created in high energy heavy-ion collisions due to both the spectators and participants are taken into account dynamically within the PHSD by including an electromagnetic tensor into the transport equations 
\cite{Voronyuk:2011jd,Toneev:2011aa,Toneev:2012zx,Voronyuk:2014rna,Toneev:2016bri,Oliva:2019kin,Oliva:2020mfr}.
Moreover, PHSD can also describe the heavy quark observables both at RHIC and LHC energies \cite{Song:2015sfa,Song:2015ykw}. In the present work, we study the impact of the electromagnetic field on heavy quark observables, mainly heavy quark directed and elliptic flows within the framework of PHSD in the presence of the electromagnetic field. 

The outline of the paper is organized as follows. In Section II  we represent the PHSD transport setup used in this calculation. In Section III we resent the electromagnetic field evolution within PHSD.
In Section IV we present the results on heavy quark directed and elliptic flows.
Finally, section V is devoted to a summary of our study.

\section{The heavy flavors production in the PHSD transport approach }

In the parton-hadron-string dynamics (PHSD) \cite{Cassing:2008sv,Cassing:2008nn,Cassing:2009vt,Bratkovskaya:2011wp,Linnyk:2015rco,Moreau:2019vhw}, the heavy flavor is produced through initial nucleon-nucleon hard scattering.
Its energy-momentum is given by the PYTHIA event generator and its spatial position by the Glauber model~\cite{Song:2015sfa,Song:2015ykw}.
The rapidity distribution and transverse momentum from PYTHIA are then rescaled such that they are consistent with  those from the FONLL  calculations~\cite{Song:2015sfa,Cacciari:2012ny}. The parton distribution in heavy-ion collisions is modified compared to  a static nucleon. As a result, heavy quark distribution, which is a by-product of the hard scattering of partons, also changes. These (anti)shadowing effects are realized through the EPS09 \cite{Eskola:2009uj} in PHSD \cite{Song:2015ykw}. After the production, heavy quarks interact with thermal partons, which are massive off-shell particles within the dynamical quasi-particle model (DQPM)~\cite{Berrehrah:2013mua}. The pole mass and spectral width of the thermal partons depend on the properties of QGP, such as temperature and baryon chemical potential~\cite{Moreau:2019vhw}. They take the form of hard thermal loop calculations and the strong coupling is parameterized such that the lattice equation-of-state (EoS) is reproduced both at zero and nonvanishing baryon chemical potentials. The scattering cross-section of heavy quarks with thermal partons is calculated by leading-order Feynman diagrams.
However, propagators in the diagram have nonzero pole mass and width, which makes the results divergence-free, and resummation of Feynman diagrams is effectively included~\cite{Berrehrah:2013mua}.

Once the local energy density is below 0.75 $\rm GeV/fm^3$, heavy quark tries coalescence for hadronization~\cite{Song:2015sfa}.
First, all possible combinations of a heavy quark and light antiquark are taken into account by calculating the coalescence probability, which depends on momentum and spatial distances between the partons in the center-of-mass frame.
In the Monte Carlo method, it is then decided whether coalescence takes place or not. If the coalescence happens, a coalescence partner is selected among all candidates by Monte Carlo based on the coalescence probability of each pair.
This process is repeated until the energy density is lower than 0.4 $\rm GeV/fm^3$.
If a heavy quark still fails coalescence, it is forced to hadronize  by the fragmentation as in pp collisions~\cite{Peterson:1982ak}. The coalescence probability is large at low transverse momentum
but small at large transverse momentum because of a poor overlap of the heavy quark and the light antiquark in momentum space.
For the heavy quark fragmentation, the PHSD adopts the Peterson fragmentation function.
After the hadronization, heavy meson interacts with light meson or light baryon.
The scattering cross sections are calculated on the basis of a chiral effective
Lagrangian with the unitarization energies \cite{Abreu:2011ic}.

It is found that the production and dynamics of heavy flavors ($D/\bar D$ and $B/\bar B$ mesons) in the PHSD is consistent with experimental observables (such as rapidity and $p_T$- spectra, $R_{AA}$ ratios, elliptic flow $v_2$ coefficients) from the RHIC beam energy scan (BES) to the large hadron collider (LHC) \cite{Song:2015sfa,Song:2015ykw,Song:2016rzw,Song:2018xca, Song:2020tfm, Song:2024hvv}.

\section{Electromagnetic fields in high-energy heavy-ion collisions}

The PHSD takes into account the dynamical formation and evolution of the electromagnetic fields (EMF) produced 
by all chargeed particles -  hadrons as well as quarks - during the time evolution of
 high-energy nucleus-nucleus and proton-nucleus  collisions \cite{Voronyuk:2011jd,Oliva:2019kin}. 
To obtain a consistent solution of quasi-particle and electromagnetic field evolution, the off-shell transport equation and the Maxwell equations for the electric field ${\bm E}$ and the magnetic field ${\bm B}$ are solved consistently.
The electric and magnetic fields can be expressed in terms of
the electromagnetic 4-vector potential $A_\mu=(\Phi,{\bf A})$:

\begin{equation}\label{fields_potentials}
{\bm E}=-\nabla\Phi-\frac{\partial{\bm A}}{\partial t}, \qquad {\bm B}=\nabla\times{\bm A}.
\end{equation}

From the Maxwell equations, we obtain the wave equation for the potentials, whose solution for arbitrarily point-like moving charges is given by the Liénard-Wiechert potentials. 
Inserting them into Eq.~\eqref{fields_potentials}, the retarded electric and magnetic fields at position ${\bm r}$, generated by a point-like source charge $e$ at position ${\bm r}'(t)$ with velocity ${\bm v}(t)$, are given by
\begin{equation}\label{Elw}
{\bm E}({\bm r},t)=\dfrac{e}{4\pi}
\left\lbrace
\dfrac{{\bm n}-{\bm\beta}}{\kappa^3\gamma^2R^2}
+\dfrac{{\bm n}\times\left[\left({\bm n}-{\bm\beta}\right)\times\dot{{\bm \beta}}\right]}{\kappa^3cR}
\right\rbrace_{\mathrm{ret}},
\end{equation}
\begin{equation}\label{Blw}
{\bm B}({\bm r},t)=\left\lbrace  {\bm n}\times{\bm E}({\bm r},t)  \right\rbrace_{\mathrm{ret}}
\end{equation}
where ${\bm R}={\bm r}-{\bm r}'$ with ${\bm r}'\equiv{\bm r}(t')$ is the relative position, ${\bm n}={\bm R}/R$ is the unit vector.  ${\bm\beta}={\bm v}/c$ and $\dot{{\bm\beta}}=\mathrm{d}{\bm\beta}/\mathrm{d}t$ are related  to the velocity and acceleration of the particle respectively, and $\kappa=1-{\bm n}\cdot{\bm\beta}$. 
All the quantities inside the braces with subscript ``ret''  have to be evaluated at times $t'$ that are solutions of the retardation equation $t'-t+{\bm R}(t')/c=0$. From Eq. ~\eqref{Elw}, we can find that retarded electromagnetic fields from moving charges are divided into two contributions. The first term  represents ``velocity fields'' which are Coulomb fields.  The second term describes ``acceleration fields'' which are interpreted as radiation fields decaying for large distances as $R^{-1}$ \cite{Landau:1975pou}.

Solving the full equation in the time-dependent case is very complicated.
If we neglect the second term ``acceleration fields'' from Eq. ~\eqref{Elw}, then 
the remaining term will be the field produced by a charge in uniform motion. Then, considering that in a nuclear collision, the total electric and magnetic field is a superposition of the fields produced from all moving charges, one can obtain the final equations implemented in the PHSD transport equation:

\begin{equation}
e{\bm E}({\bm r},t)=\sum_i
\dfrac{\mathrm{sgn}(q_i)\alpha_{em}{\bm R}_i(t)(1-\beta_i^2)}{\left\lbrace\left[{\bm R}_i(t)\cdot{\bm\beta}_i\right]^2+R_i(t)^2\left(1-\beta_i^2\right)\right\rbrace^{3/2}},
\end{equation}
\begin{equation}
e{\bm B}({\bm r},t)=\sum_i
\dfrac{\mathrm{sgn}(q_i)\alpha_{em}{\bm\beta_i}\times{\bm R}_i(t)(1-\beta_i^2)}{\left\lbrace\left[{\bm R}_i(t)\cdot{\bm\beta}_i\right]^2+R_i(t)^2\left(1-\beta_i^2\right)\right\rbrace^{3/2}},
\end{equation}
where $\alpha_{em}=e^2/4\pi\simeq1/137$ is the electromagnetic fine-structure constant, and the summation of $i$ runs over all charge particles with charge $q_i$. 
The quasiparticle propagation in the electromagnetic field is calculated by the Lorentz force:
\begin{equation}\label{lorentz}
\left(\dfrac{\mathrm{d}{\bm p}_i}{\mathrm{d}t}\right)_{em}=q_i\left({\bm E}+{\bm\beta}_i\times{\bm B}\right).
\end{equation}

For the details of the electromagnetic filed production and propagation in high-energy heavy ion collisions within PHSD, we refer to 
\cite{Voronyuk:2011jd,Toneev:2011aa,Toneev:2012zx,Voronyuk:2014rna,Toneev:2016bri,Oliva:2019kin,Oliva:2020mfr}.
It is important to mention that the time evolution of the electromagnetic fields computed within PHSD accounts naturally for the electric conductivity $\sigma_{el}$ of the system.  On the other side,  $\sigma_{el}$ can be computed in the PHSD as the response of the strongly interacting system in equilibrium to an external electric field  \cite{Cassing:2013iz}.
We note that $\sigma_{el}$ in the PHSD  (due to the DQPM) is temperature-dependent \cite{Soloveva:2019xph, Fotakis:2021diq}, i.e. the ratio $\sigma_{el}/T$ is rising with $T$ in line with lQCD data \cite{Brandt:2012jc, Aarts:2014nba}.

\section{Results}

Within the PHSD transport approach  we proceed to evaluate the heavy quark directed flow,
\be
 v_1=\left\langle \cos(\phi) \right\rangle =\left\langle  \frac{p_x}{p_T}\right\rangle\ , \qquad \qquad
\ee
considered as a novel observable to probe the EMF produced in high-energy nucleus-nucleus collisions, where $\phi$ is the azimuthal angle with respect to the reaction plane.  We will also compute the  $D$ meson elliptic flow, 
\be
 v_2=\left\langle \cos(2\phi) \right\rangle =\left\langle  \frac{p_x^2 -p_y^2}{p_x^2+p_y^2}\right\rangle\ , \qquad \qquad
\ee
a measure of the anisotropy in the angular distribution of the $D$ meson, to study the influence of EMF on the elliptic flow.

\begin{figure}[t!]
	\begin{center}
		\includegraphics[scale=0.32]{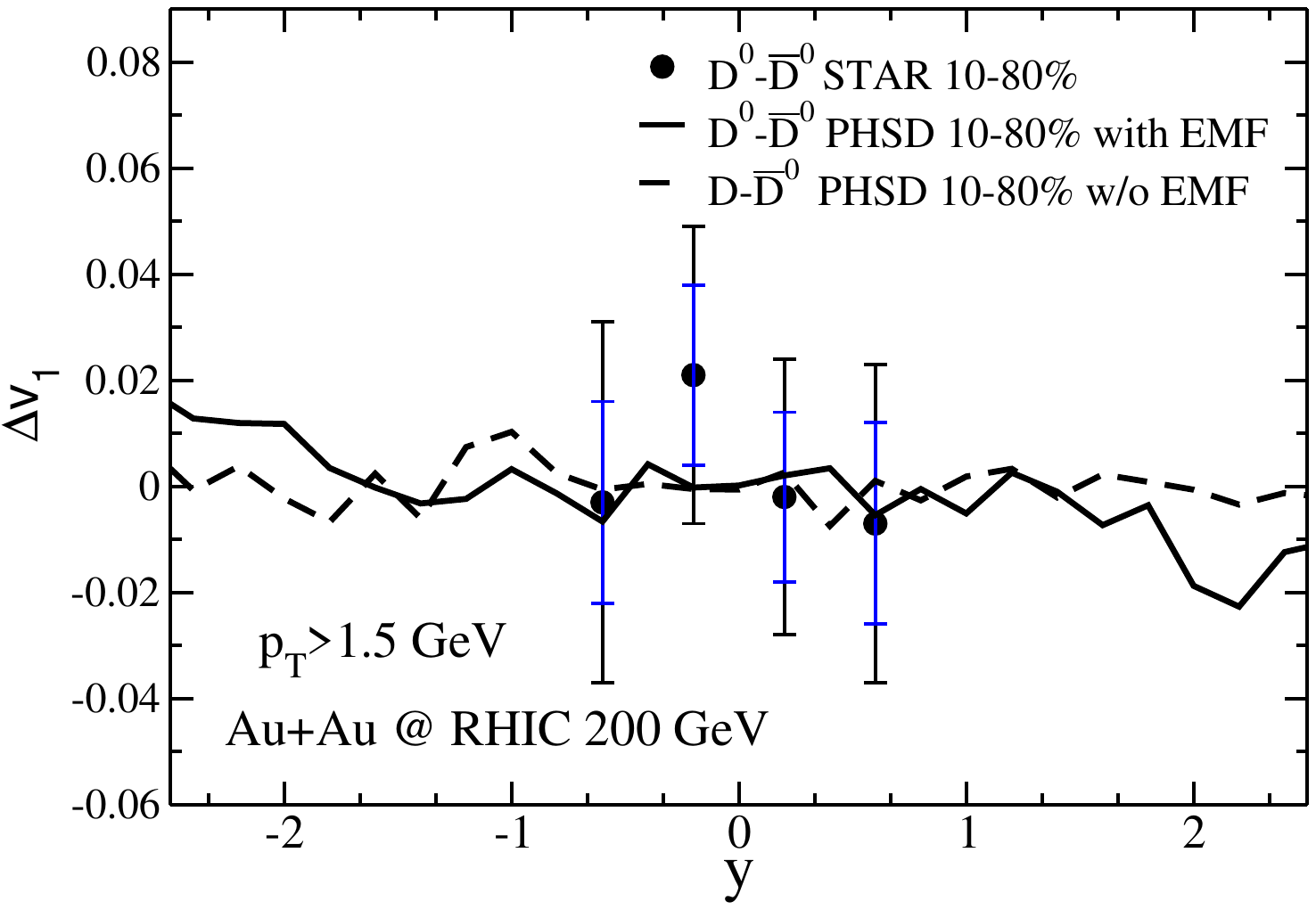}
	\end{center}
	\caption{\label{Fig:v1}Variation of the directed flow splitting, $\Delta v_1$ as a function of rapidity for 10--80\% central Au+Au collisions at the highest RHIC energy $\sqrt{s_{NN}}=200$ GeV. The experimental data of the STAR Collaboration are taken from Ref.~\cite{STAR:2019clv}.}.
\end{figure}

In Fig.~\ref{Fig:v1} we show the variation of the directed flow  splitting, $\Delta v_1=v_1(D^0)-v_1(\overline{D^0})$, as a function of rapidity for the highest RHIC energy in comparison with the STAR data ~\cite{STAR:2019clv}. We have computed the splitting to exclude the contribution to the directed flow from the bulk evolution and to highlight only the effect coming from the EMF.  $D^0$ and $\overline{D^0}$, being neutrals, are produced by hadronization of $c$ and $\overline{c}$ quarks respectively. Hence, the 
$\Delta v_1$ of the $D^0$ and $\overline{D^0}$ mesons, if any, is driven by the  $c$ and $\overline{c}$ quarks. We find a very mild effect of the EMF on the $D$ meson directed flow, mainly at larger rapidity, which is consistent with the available STAR data. Within the current accuracy, it is almost zero in the rapidity range explored in the STAR measurement. However, at large backward rapidity, $\Delta v_1(y)$ gets a positive contribution from the electromagnetic fields and a  negative contribution at larger forward rapidity. We obtain a negative 
slope of the $\Delta v_1$ at the highest RHIC energy. However, the magnitude of the splitting obtained within PHSD is less in comparison with the results presented in Ref.~\cite{Oliva:2020doe}. This is mainly due to the different EMF computations.

\begin{figure}[t!]
	\begin{center}
		\includegraphics[scale=0.32]{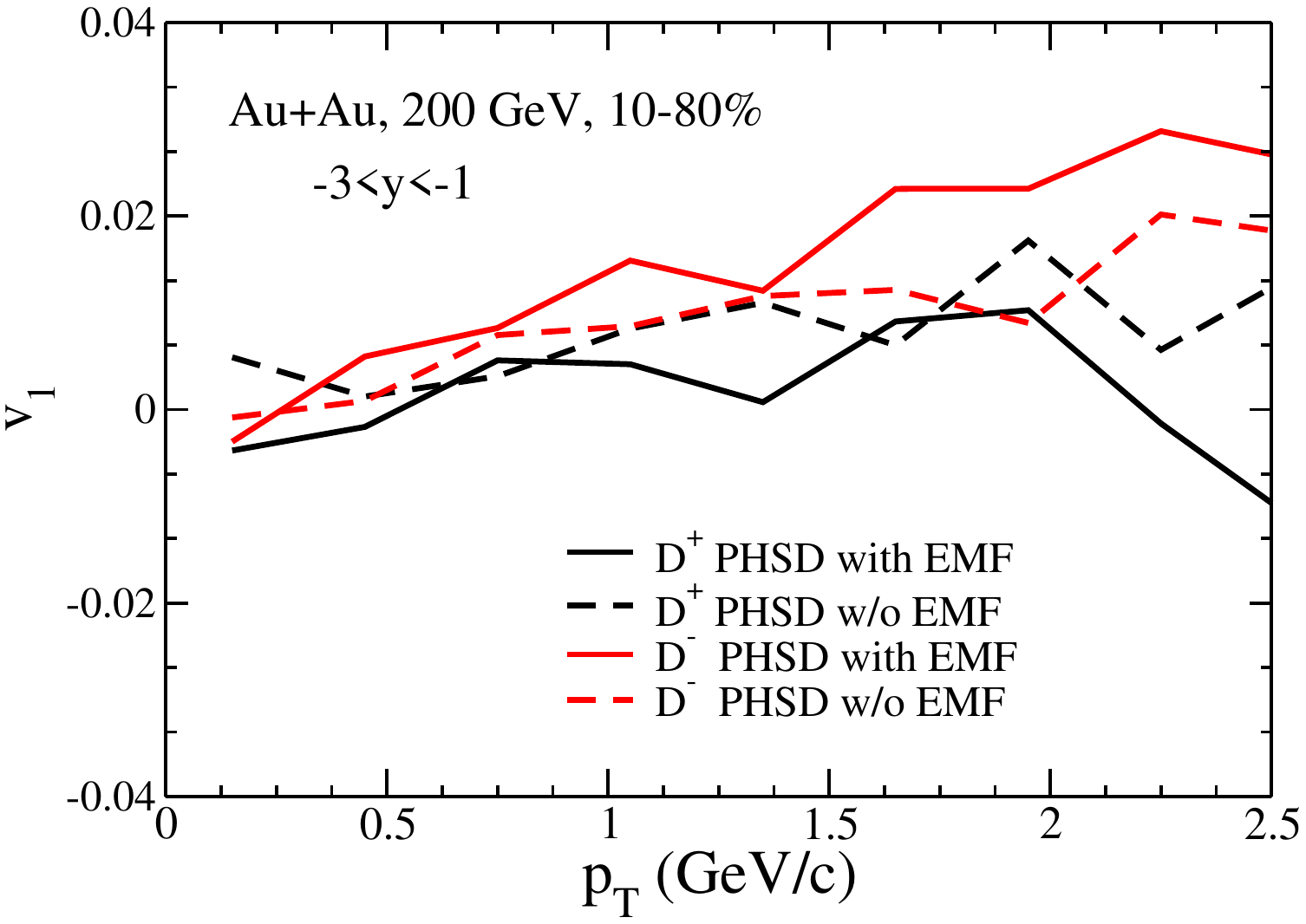}
	\end{center}
	\caption{\label{Fig:v1pt} Directed flow $v_1$ of $D^+$ and $D^-$ mesons  as a function of $p_T$ for 0--80\% central Au+Au collisions at $\sqrt{s_{NN}}$=200 GeV for $-3 < y < -1$.}
\end{figure}

In Fig.~\ref{Fig:v1pt} we show the directed flow, $v_1$, of $D^+$ and $D^-$ as a function $p_T$ in the backward rapidity ($-3\le y \le -1)$ with and without EMF.  $D^+$ and $D^-$ mesons are produced by hadronization of $c$ and $\overline{c}$ quarks respectively, like the $D^0$ and $\overline{D^0}$ mesons. We observe a splitting in the $v_1$ of $D^+$ and $D^-$ mesons due to the presence of the EMF in the given rapidity window. We find that the $v_1$ splitting - as a function of $p_T$ - is non-zero and quite substantial contrary to  the $v_1$ splitting as a function of rapidity. The splitting of the $D$ meson directed flow as a function of $p_T$, if measured in  future experiment, can act as a novel probe of the produced EMF.   

\begin{figure}[t!]
	\begin{center}
		\includegraphics[scale=0.32]{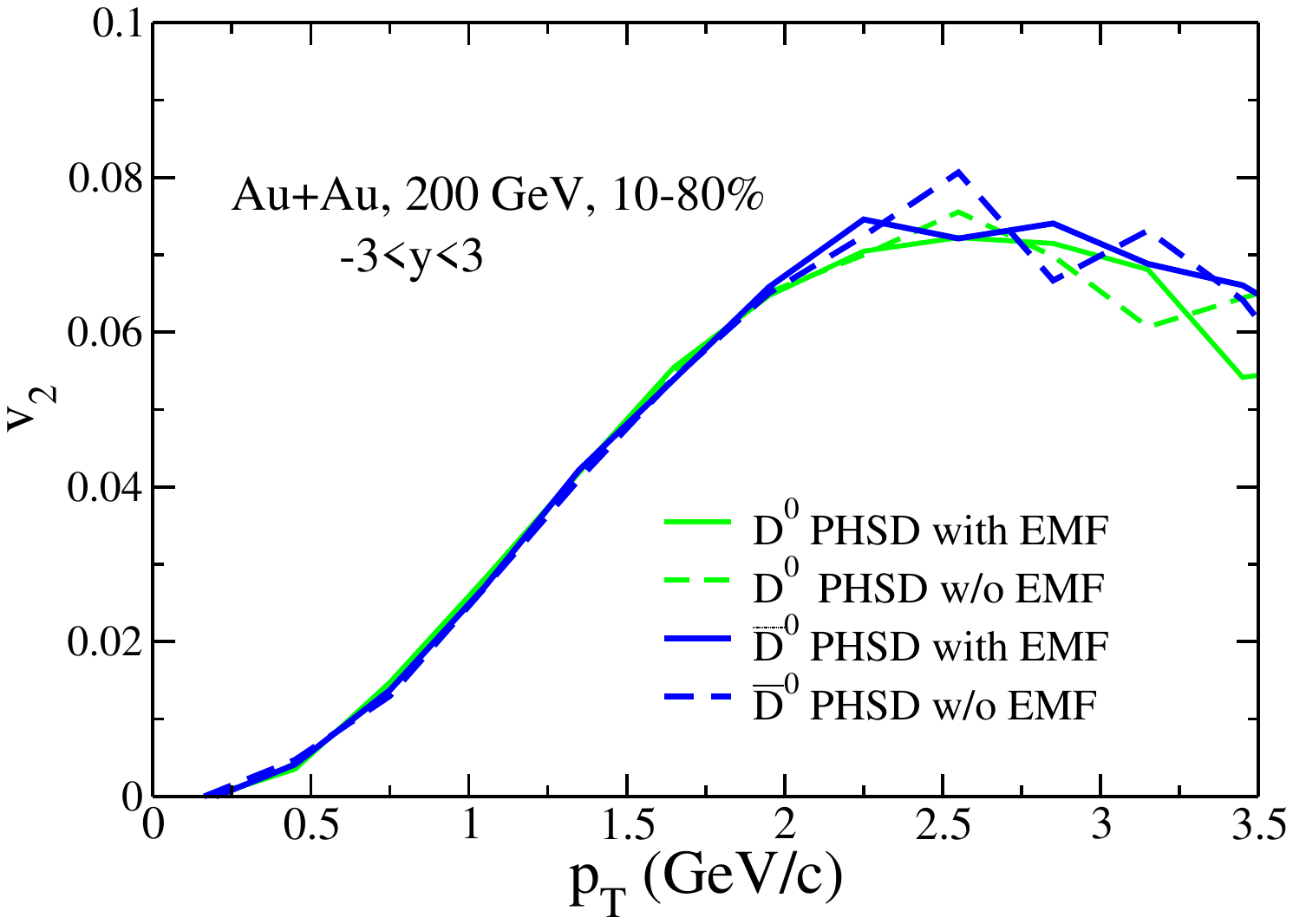}
	\end{center}
	\caption{\label{Fig:v2}Elliptic flow, $v_2$, of $D^0$ and $\overline{D}^0$ mesons  as a function $p_T$ for 10--80\% central Au+Au collisions at the highest RHIC energy  $\sqrt{s_{NN}}$=200 GeV.}
\end{figure}

We have also computed the $D$ meson elliptic flow, $v_2$, in the PHSD transport approach with and without EMF to highlight the possible effect of the EMF on $D$ meson elliptic flow. In Fig.~\ref{Fig:v2} we present the  elliptic flow of $D^0$ and $\overline{D^0}$ as a 
function of $p_T$ with and without EMF. We observe that the impact of the EMF on the $D$ meson elliptic flow is negligible.

\begin{figure}[t!]
	\begin{center}
		\includegraphics[scale=0.3]{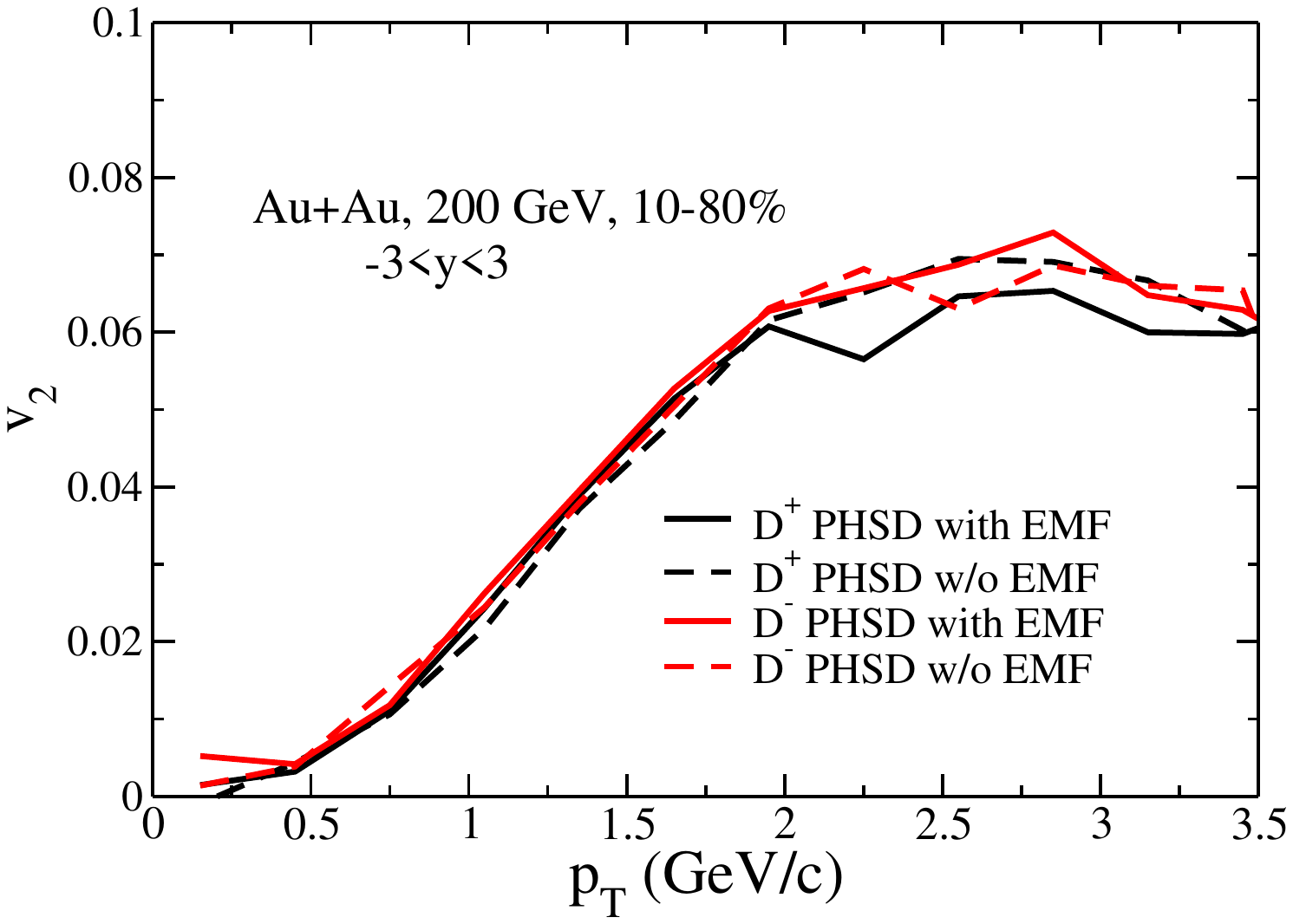}
	\end{center}
	\caption{\label{Fig:v2_1}Elliptic flow, $v_2$, of $D^+$ (upper) and $D^-$ mesons (lower) as a function $p_T$ for 10--80\% central Au+Au collisions at $\sqrt{s_{NN}}$=200 GeV.}
\end{figure}

 In Fig.~\ref{Fig:v2_1} we present the variation of the elliptic flow of $D^+$ and $D^-$ as a function of $p_T$ with and without EMF. We find that the impact of the EMF of the $D$ meson elliptic flow is small.

\section{Summary and outlook} 

We have studied the dynamics of charm quarks in the QCD matter produced in nucleus-nucleus collisions at the highest RHIC energy, taking into account the impact of electromagnetic fields within the PHSD transport approach, a microscopic covariant dynamical approach for strongly interacting matter formulated on the basis of off-shell Kadanoff-Baym equations. 
PHSD can describe heavy quark observables, the nuclear suppression factor ($R_{AA}$), and elliptic flow ($v_2$), both at RHIC and LHC energies. In the present study, the initial EMF created in high energy heavy-ion collisions due to both the spectators and participants is taken into account dynamically  within PHSD, the off-shell transport equation, and the Maxwell equations for the EMF are solved  to obtain a consistent solution of quasi-particle and electromagnetic field evolution.
The electromagnetic fields are calculated according to the retarded Lienard-Wiechert equations for a charge moving with a certain velocity, then summed over all the charged quasi-particles in the medium, both participants and spectators.  The time evolution of the EMF computed within PHSD naturally accounts for the electric conductivity $\sigma_{el}$ of the system that is temperature dependent and in line with the lattice results.

We would like to recall that the measured slope of the $D$ meson directed flow splitting both at RHIC and LHC is a subject of high contemporary interest. All the existing model calculations can not describe the slope simultaneously for both RHIC and LHC energies. In this present calculation, we have made an attempt to study heavy quark dynamics in the background EMF at RHIC energy in a self-consistent way, hence, relaxing some of the approximations made in earlier studies by considering only spectators to 
evaluate the EMF including a constant electric conductivity $\sigma_{el}$.

We have computed the splitting of the directed flow, $\Delta v_1=v_1(D^0)-v_1(\overline{D^0})$, within the PHSD transport approach and compare the results with the available STAR data at the highest RHIC energy. PHSD can describe the STAR data, however,  the impact of the EMF is quite mild, and it is almost zero in the rapidity ranges explored in the STAR measurement. However, the impact of the EMF for large forward-backward rapidity is quite visible. We obtain a negative slope of the $\Delta v_1$ at the highest RHIC energy, though the magnitude of the splitting obtained within PHSD is small 
in comparison with other models. We have also evaluated the $D$ meson directed flow as a function of $p_T$. 

We have observed that the $D$ meson directed flow splitting as a function of transverse momentum at a certain rapidity range is very sensitive to the EMF. If measured in an experiment, it can act as a novel probe to characterize the produced the EMF at high-energy heavy-ion collisions. We have also computed the possible impact of EMF on $D$ meson elliptic flow. 

We have found that the impact of the EMF on the $D$ meson elliptic flow is negligible. This indicates  that the heavy quark directed flow is the only observable to characterize the EMF. However, the directed flow as a function of $p_T$ is more sensitive to the electromagnetic filed. It will be interesting  to perform a similar study in the presence of an electromagnetic field  at LHC energy. We will address this in a forthcoming article.

\begin{acknowledgements}
S.K.D. acknowledges the support from SERB IRE fellowship, India, having Project No. SIR/2022/000231. 
The authors thank to Lucia Oliva for the inspiring discussions. 
We also acknowledge the support by the Deutsche Forschungsgemeinschaft (DFG) through the grant CRC-TR 211 "Strong-interaction matter under extreme conditions" (Project number 315477589 - TRR 211). 
\end{acknowledgements}

\bibliography{biblio_HQ_EM}

\end{document}